\documentclass[10pt,english]{article}
\usepackage{subfigure}
\usepackage{graphicx}
\usepackage{float}
\usepackage[T1]{fontenc}
\usepackage[latin9]{inputenc}
\usepackage[margin=1.5in]{geometry}
\usepackage{float}
\usepackage{amsmath}
\usepackage{amsbsy}
\usepackage{setspace}
\usepackage{amssymb}
\usepackage{esint}
\usepackage{cite}
\newfloat{algorithm}{tbp}{loa}
\floatname{algorithm}{Algorithm}
\usepackage{algorithmic}
\usepackage{hyperref}

\newlength\myindent
\makeatletter

\floatstyle{ruled}
\newfloat{algorithm}{tbp}{loa}
\floatname{algorithm}{Algorithm}

\usepackage{babel}

\begin{document}

\title{Growing Random Strings in CA}

\author{M. Andrecut}


\maketitle
{

\centering Calgary, Alberta, T3G 5Y8, Canada

\centering mircea.andrecut@gmail.com

} 

\bigskip 

\begin{abstract}

We discuss a class of cellular automata (CA) able to produce long random strings, starting from short "seed" strings. 
The approach uses two principles borrowed from cryptography: diffusion and confusion. We show numerically that the 
strings are pseudo-random using three approaches based on: Fourier transform, entropy estimation, and compression. 
An application to cryptography is also included with the corresponding Python code. 

\bigskip


\end{abstract}

\bigskip

\section{Introduction}

Cellular Automata (CA) are discrete time and space dynamical systems, consisting of an array of identical cells, each cell implementing a finite state automaton. 
Formally, a CA is a 4-tuple $\{\mathcal{C}, \mathcal{S}, \mathcal{N}, \mathcal{R}\}$, where: $\mathcal{C}$ is the cellular space;
$\mathcal{S}$ is the finite set of states of each cell; $\mathcal{N}$ is the neighborhood of a cell; and $\mathcal{R}$ is the local rule used to compute a cell's next state.
The cellular space can be $d$-dimensional, and the neighborhood of a cell can have various configurations. Each cell uses the state of its neighbors and its own state to 
compute the next state. While some CA can have very simple architectures and rules, they still display surprisingly complex behavior \cite{key-1}.

Due to their complex chaotic behavior, CA have been proposed as random number generators in cryptography. 
This direction was initiated by Stephen Wolfram's claims about rule 30 implemented in an elementary CA \cite{key-2}. 
This random generator passed a standard suite of statistical randomness tests, and its output was used 
as a secret keystream in symmetric encryption. 

More recently, one-dimensional CA have been shown to be isomorphic to Linear Feedback Shift Registers (LFSRs) \cite{key-3}, which are probably the most used 
systems for the generation of pseudorandom sequences with applications to cryptography and communications \cite{key-4}. Thus, CA can be considered as 
an viable alternative for such applications \cite{key-5}. 

Typical CA are one-dimensional fixed length registers with $L$-cells, where the cell states are updated synchronously according to the local transition rule. 
Contrary to this, here we consider a linearly growing class of CA, able to produce long random strings, starting from short "seed" strings. 
The local transition rule is governed by two principles borrowed from cryptography: diffusion and confusion \cite{key-6}, \cite{key-7}. We show numerically that the 
resulted strings are pseudo-random using three approaches based on: Fourier transform, entropy estimation, and compression. 
An application to cryptography is also included with the corresponding Python code

\section{Growing random strings in CA}

We assume that at time $t$ the CA is represented as a cell register $\mathcal{C}(t)$ of length $L(t)$ depending on time $t$, where each cell is one byte (8-bits), such that the maximum number of states is $K=256$, and $\mathcal{S}=\{0,1,2,...,255\}$. 
The state of the CA at time $t$ is therefore: 
\begin{equation}
\mathcal{C}(t) = [s_0(t),s_1(t),...,s_{L(t)-1}(t)], \quad s_\ell(t) \in \mathcal{S}, 0\leq \ell \leq L(t). 
\end{equation}
The initial state at $t=0$ is:
\begin{equation}
\mathcal{C}(0) = [s_0(0),s_1(0),...,s_{L(0)-1}(t)], \quad s_\ell(0) \in \mathcal{S}, 0\leq \ell \leq L(t), 
\end{equation}
where $L(0) \geq L^* > 0$ is the initial length, and $L^*$ is a minimum accepted length.
Using the "diffusion" (or mixing) principle, at each time $t$ we apply the following local rule:
\begin{equation}
s_\ell(t+1) = (s_{(\ell-2) \text{mod} L(t)}(t)+s_{(\ell-1) \text{mod} L(t)}(t)+s_{\ell}(t))  \text{mod} K, \quad \ell = 0,1,...,L(t)-1
\end{equation}
Each "diffusion" is followed by a "confusion" step, where the length of the CA is increased by appending a new cell having the state:
\begin{equation}
s_{L(t+1)}(t+1) = s_{L(t)+1}(t+1) = \left( \sum_{n = 0}^{L(t)-1} s_n(t) \right)  \text{mod} K,
\end{equation}
such that:
\begin{equation}
L(t+1) = L(t) + 1.
\end{equation}
Thus, at each iteration step the length of CA grows with one cell.

\begin{figure*}[h!]
\centering {\includegraphics[width=13cm]{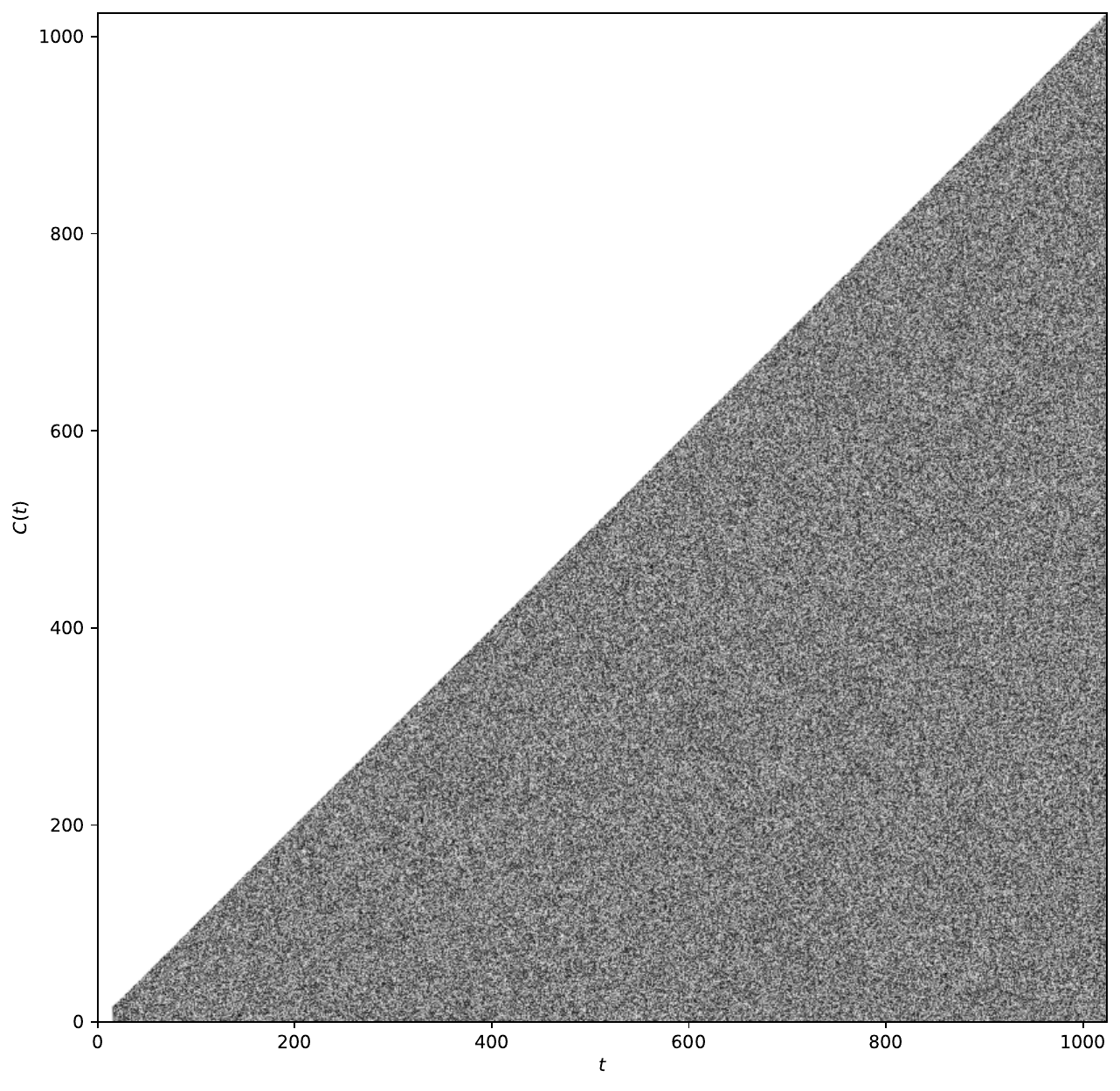}}
\caption{The CA growth from $L(0)=2^4$ to $L(T)=2^{10}$.}
\end{figure*}

In Figure 1 we show the time evolution of the CA growth, for $L(0)=2^4$ and $t=2^4,...,T$, where $T=2^{10}$. One can see that while the length grows from 
$L(0)=2^4$ to $L(T)=2^{10}$ the CA exhibits random behavior. 

\begin{figure*}[h!]
\centering {\includegraphics[width=9cm]{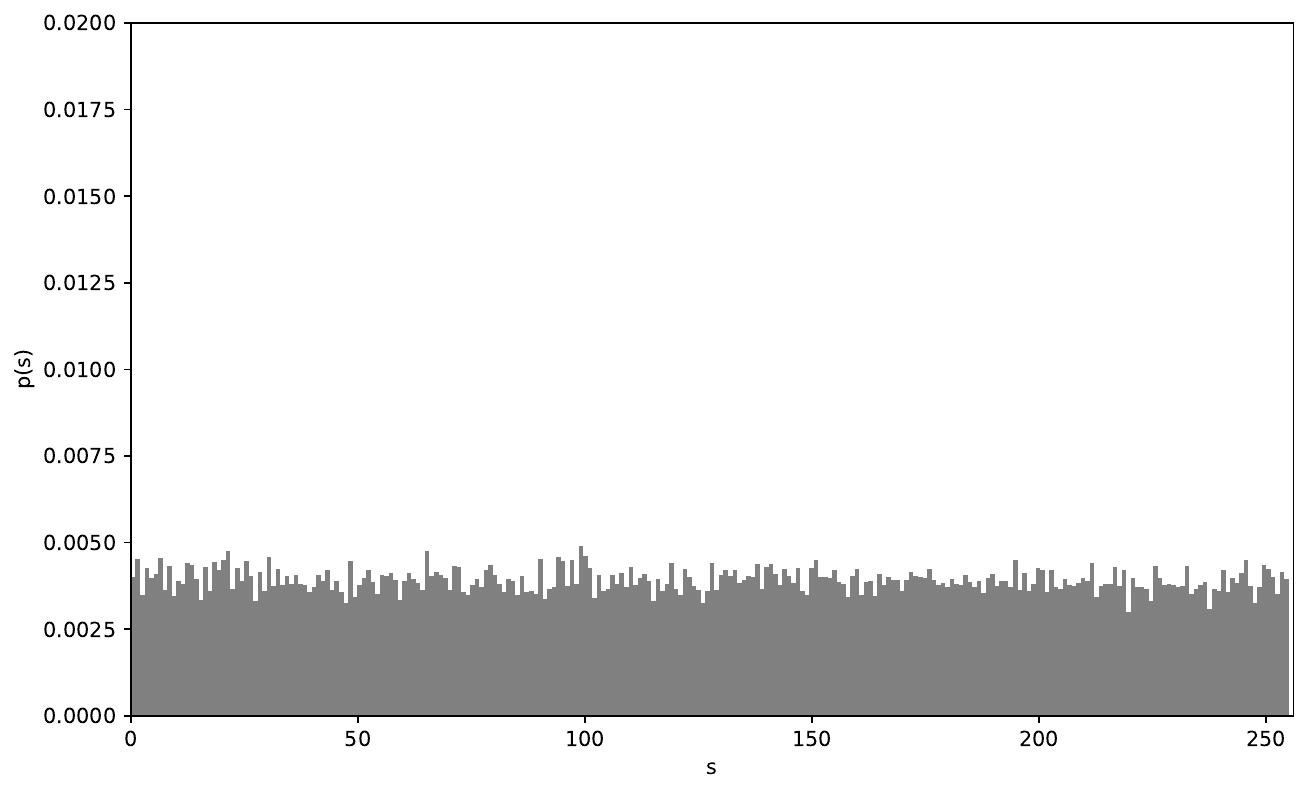}}
\caption{The the probability distribution of states, for $L=2^{15}$.}
\end{figure*}

The main question is: how to prove that the resulted strings are pseudo-random? 
A first option would be to show that the probability distribution of states is uniform. In Figure 2 we show the probability distribution for $L=2^{15}$, which is clearly uniform.

\begin{figure}[h!]
\centering
\subfigure{\includegraphics[width=8.5cm]{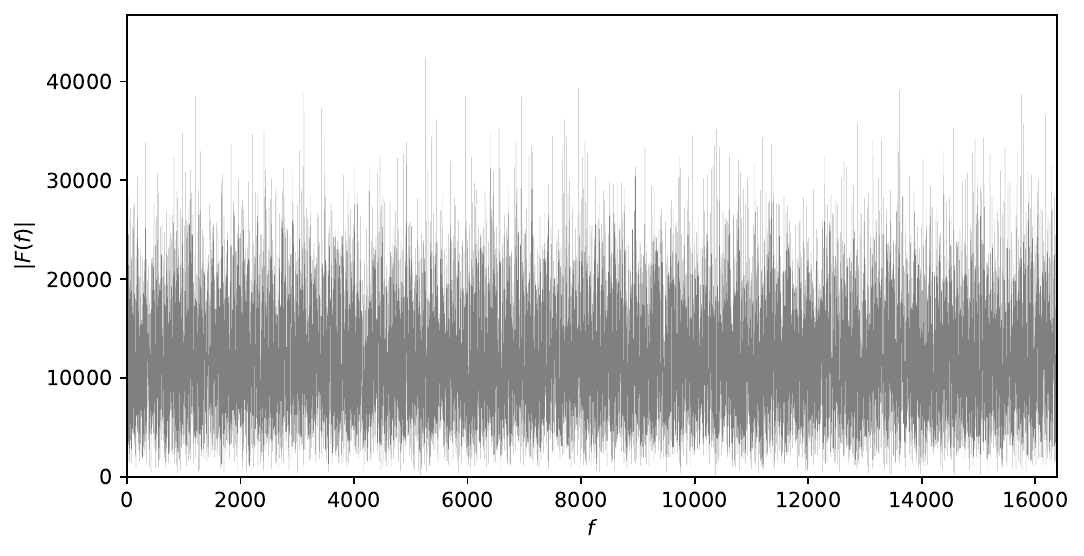}}
\subfigure{\includegraphics[width=7.5cm]{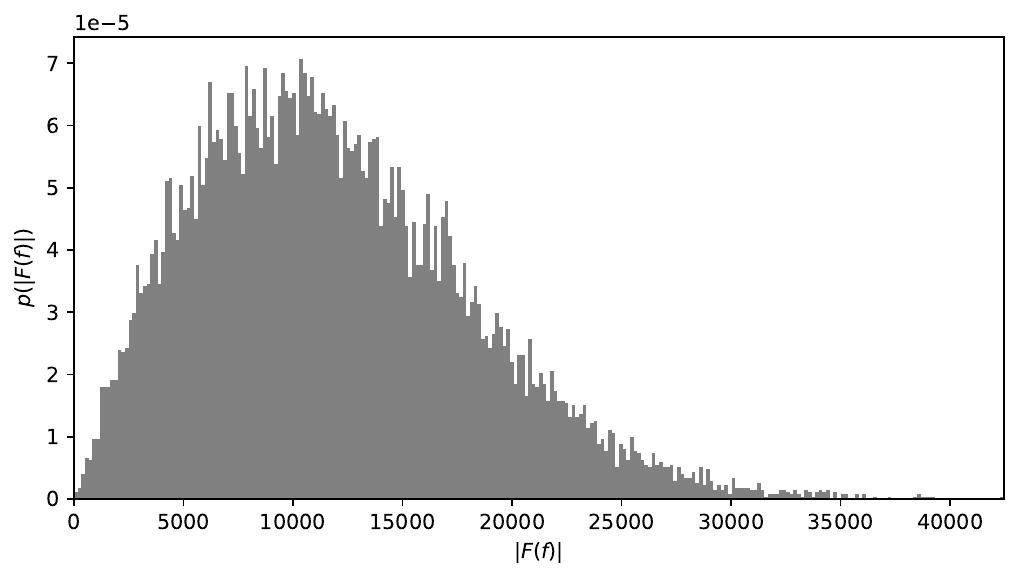}}
\subfigure{\includegraphics[width=7.75cm]{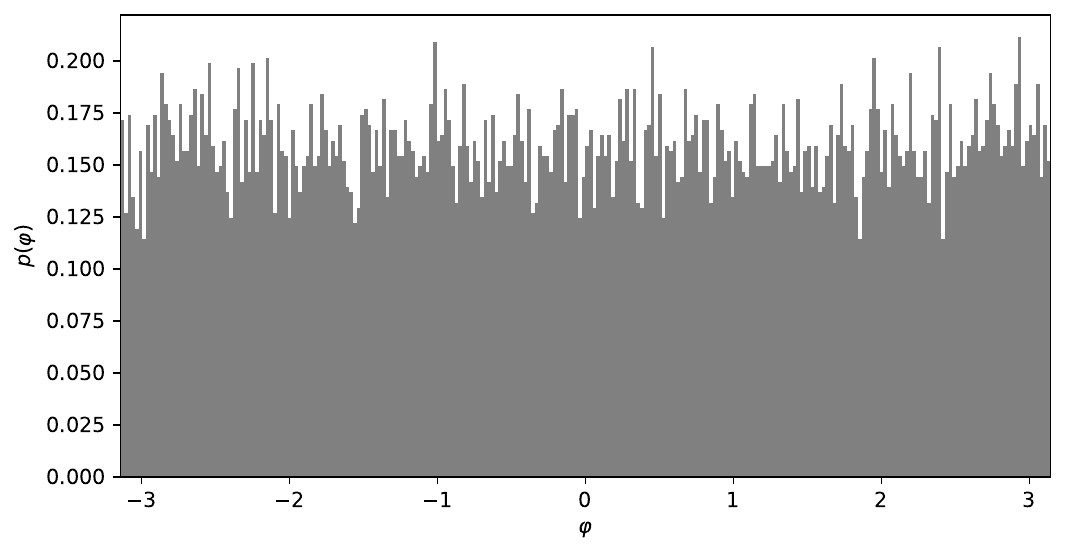}}
\caption{The Fourier analysis for $L=T=2^{15}$: (top) amplitude coefficients; (middle) amplitude coefficients distribution; (bottom) phase angle distribution.}
\end{figure}

Another option would be to use the fast Fourier transform of the last state of the CA, $F(f) = FFT(C(T))$. In Figure 3 we give the Fourier analysis for $L=T=2^{15}$. The amplitude of the coefficients 
seems uniform, and doesn't show periodicity. The amplitude coefficients show an expected Rayleigh distribution:
\begin{equation}
p(x) = x\sigma^{-2} e^{-x^2\sigma^{-2}/2}, \quad x \geq 0,
\end{equation}
which is the probability distribution for nonnegative random variables, 
while the distribution of the Fourier transform phase is uniform. 

Two important measures of the randomness are the Shannon entropy \cite{key-8}:
\begin{equation}
H = - \sum_{k= 0}^{K-1} p_k \log_K p_k,
\end{equation}
and the compression ratio. Here, $p_k$ is the probability (frequency) of the state $k\in \mathcal{S}$. The compression rate 
can be estimated using the BZIP2 algorithm \cite{key-9} as following:
\begin{equation}
r = \text{length}(BZIP2(\mathcal{C}(T))/\text{length}(\mathcal{C}(T)),
\end{equation}
where $BZIP2(\mathcal{C}(T))$ is the compressed value of the last CA state $\mathcal{C}(T)$. In our case, for a single run, 
the computed entropy is $H=0.9993882799247457$, and the compression ratio is $r=1.073150634765625$. Which means 
that the entropy is very close to the maximum value $H_{max}=1$, and the BZIP2 could not actually compress the state of the CA, 
which means it is pseudo-random. The Python implementation of the above analysis is given below:

\footnotesize
\begin{verbatim}
import math
import bz2
import lzma
import numpy as np
from collections import Counter
import matplotlib.pyplot as plt

def runCA(C,L):
    C,M,Q = bytearray(C.encode()),len(C), np.zeros((L,L))
    for J in range(M,L):
        C = [(C[n-2] + C[n-1] + C[n]) % 256 for n in range(J)]
        C.append(sum(C) % 256)
        Q[J,0:J+1] = C
    return 255-Q

def entropy(C):
    p, lns = Counter(C), float(len(C))
    return -sum(count/lns * math.log(count/lns, 256) for count in p.values())

if __name__== "__main__":
    C = "abcdefghijklmnop"
    C = runCA(C,32768)
    F = np.fft.fft(C[-1]-np.mean(C[-1]))[:len(C)//2]

    fig = plt.figure(figsize=(10,10))
    plt.imshow(C[0:1024,0:1024].T, cmap='gray',interpolation=None)
    plt.xlabel('t'); plt.ylabel('C(t)')
    plt.xlim([0, 1024]); plt.ylim([0, 1024])
    fig.savefig("fig1.pdf",bbox_inches="tight")

    fig = plt.figure(figsize=(10,6))
    plt.hist(C[-1],256, facecolor='gray', density=True)
    plt.xlim([0, 256]);plt.ylim([0, 0.02])
    plt.xlabel('s'); plt.ylabel('p(s)')
    fig.savefig("fig2.pdf",bbox_inches="tight")

    fig = plt.figure(figsize=(8,4))
    plt.plot(np.arange(len(F)),np.abs(F),linewidth=0.1,color="gray")
    plt.xlim([0, len(F)]); plt.ylim([0, 1.1*np.max(np.abs(F))])
    plt.xlabel('$f$'); plt.ylabel('$|F(f)|$')
    fig.savefig("fig3a.pdf",bbox_inches="tight")

    fig = plt.figure(figsize=(8,4))
    plt.hist(np.abs(F),256, facecolor="gray", density=True)
    plt.xlim([0, np.max(np.abs(F))])
    plt.xlabel('$|F(f)|$'); plt.ylabel('$p(|F(f)|)$')
    fig.savefig("fig3b.pdf",bbox_inches="tight")

    fig = plt.figure(figsize=(8,4))
    plt.hist(np.angle(F),256, facecolor="gray", density=True)
    plt.xlim([np.min(np.angle(F)),np.max(np.angle(F))])
    plt.xlabel(r'$\varphi$'); plt.ylabel(r'$p(\varphi)$')
    fig.savefig("fig3c.pdf",bbox_inches="tight")

    H = entropy(bytes(C[-1]))
    print("Entropy = ", H)
    Z = bz2.compress(bytes(C[-1]))
    print("BZIP2 compression ratio = ", len(Z)/len(C[-1]))
\end{verbatim}
\normalsize

\section{Application to cryptography}

Ciphers are widely used cryptographic algorithms in confidential communication systems. 
In such a system, the sender is encrypting the "plaintext" with an algorithm (cipher) and a secret key. 
The resulted encrypted information is called "ciphertext" and is sent to the receiver, which later 
can recover the plaintext by using a secret key and the corresponding decryption algorithm. If the secret keys used for 
encryption and decryption are the same then the process is called "symmetric encryption", and requires the 
sender and the receiver to share the secret key over a secure channel which cannot be eavesdropped (Fig. 4). 

\begin{figure*}[h!]
\centering {\includegraphics[width=11cm]{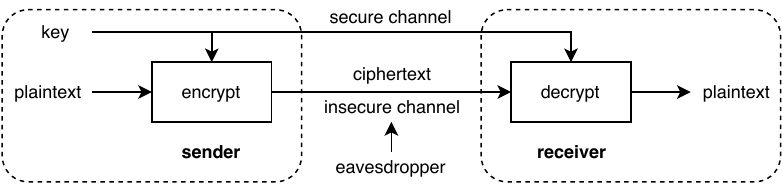}}
\caption{The symmetric encryption protocol.}
\end{figure*}

Let us denote by $\mathcal{X}$, $\mathcal{K}$, $\mathcal{Y}$ the sets of all messages, keys and respectively ciphertexts. 
We also assume that $E(k,x) = y$ and $D(k,y) = x$, are the encryption and respectively decryption functions, with $x\in \mathcal{X}$, $k\in \mathcal{K}$, and respectively $y\in \mathcal{Y}$.
In order to define a cipher, the $E$ and $D$ functions must satisfy the following correctness property $D(k,E(k,x)) = x, \forall x \in \mathcal{X}$, which means that $D$  must be the inverse of $E$, $D=E^{-1}$.

A "perfect cipher" does not provide an attacker with any additional information about the plaintext, given the ciphertext \cite{key-1,key-2}. 
This means that for any two messages $x,x^*\in \mathcal{X}$, the probability that $x=x^*$ is $|\mathcal{X}|^{-1}$,
where $|\mathcal{X}|$ is the size (cardinal) of $\mathcal{X}$.

The main idea we want to exploit here is that the growing CA described in the previous section can be "seeded" with the secret key and then its state can grow into 
a random state $k$ that then can be used to encrypt the plain text $x$ using the Vernam cipher: $y_n = k_n \oplus x_n$, $n=0,1,...,N-1$, 
where $N$ is the length of the plain text, and $\oplus$ is the XOR bitwise operation. The decryption is symmetrical, one can 
recover the plain text using the same key, by applying again the XOR operation: $x_n = k_n \oplus y_n$. 

The Python code implementing the class vlCryptography.py is straightforward, and it is given below:

\footnotesize
\begin{verbatim}
class vlCryptography(object):
    def __init__(self, key):
        if len(key) < 9: raise ValueError("Key length must be minimum 9 characters.")
        self._key = key.encode()

    def crypt(self,x):
        if len(x) == 0: raise ValueError("Empty source.")
        key,x,M,N = bytearray(self._key),bytearray(x),len(self._key),len(x)
        for J in range(M,N):
            key = [(key[m-2] + key[m-1] + key[m]) % 256 for m in range(J)]
            key.append(sum(key) % 256)
        return bytes([key[n]^x[n] for n in range(N)])
\end{verbatim}
\normalsize

Here, "vl" in vlCryptography stands for "very light", because the class doesn't require any other particular Python module. 
An usage example is the following:
\footnotesize
\begin{verbatim}
import lorem
import base64
from vlCryptography import vlCryptography

key = lorem.sentence()
print("Key:")
print(key,"\n")

vlC = vlCryptography(key)

plain_text = lorem.paragraph()
print("Plain text:")
print(plain_text,"\n")

cipher_text = vlC.crypt(plain_text.encode())
print("Cipher text:")
print(base64.urlsafe_b64encode(cypher_text).decode(),"\n")

print("Plain text:")
plain_text = vlC.crypt(cipher_text)
print(plain_text.decode())
\end{verbatim}
\normalsize

For testing purposes, here we use the "lorem" module, which is text generator that looks like Latin. Also, the cipher text is encoded in base64, such that it can be printed. 
The first step is to generate a secret key using the "lorem" sentence method, then an instance of the vlCryptography object is initialized using the key, 
the plain text is generated using the "lorem" paragraph method, and then is encrypted using the "crypt" method of the vlCryptography class. Finally, the plain text 
is recovered from the cipher text using the same crypt method. Below we give an output example:
\footnotesize
\begin{verbatim}
Key:
Est aliquam velit sed. 

Plain text:
Est labore neque adipisci dolorem ut labore sit. Magnam etincidunt aliquam porro dolorem 
neque modi ipsum. Quaerat etincidunt quiquia consectetur. Numquam quiquia ipsum modi 
quiquia amet numquam sed. Amet tempora adipisci voluptatem.  

Cipher text:
dHESbMrGrk8ZT4C7IRnsRdFOZjroJiyaBpOeV6TQeAvO8pOU4NVx6qksIqyHaYR83sDl6Ts_3e-wxPFyEPZND-BO
ff0koi1iwJm7zrHBwucBxBXtdHtFZXoLiCkOkbq5yxIQiQSp46zg6dZaDLsEesNrz-mOxKqlNTnkJCnHhCAQ-xW-
1AmhoWKAEyPKyFP7WM9pOgz3r4v670ABBBvTQ4BPINBuMnAzHhSaLnpf-ex6AdL2bArschzmn8kLFcsTADKefasC
VfpntSPVX0DyY_evxpwqqjajqA5FejN_loU-ae-9k5eQ4y4= 

Plain text:
Est labore neque adipisci dolorem ut labore sit. Magnam etincidunt aliquam porro dolorem 
neque modi ipsum. Quaerat etincidunt quiquia consectetur. Numquam quiquia ipsum modi 
quiquia amet numquam sed. Amet tempora adipisci voluptatem.
\end{verbatim}
\normalsize

\section{Conclusion}
Typical CA are characterized by fixed length registers. Contrary to this, here we have considered a linearly growing class of CA, 
able to produce long random strings, starting from short "seed" strings. 
The local transition rule of the CA is governed by two principles borrowed from cryptography: diffusion and confusion. 
We have shown numerically that the resulted strings are pseudo-random using three approaches based on: Fourier transform, entropy estimation, and compression. 
An application to cryptography, and the corresponding Python code, was also included. 

The code is available at: \url{https://github.com/mandrecut/vlCryptography}.

\end{document}